\documentstyle[preprint,aps]{revtex}
\draft
\begin{document}
\title{Alkali Bose Condensate Mixtures} 

\author{Tin-Lun Ho and V.B. Shenoy}
\address{Physics Department,  The Ohio State University, Columbus, Ohio
43210}
\maketitle

\begin{abstract}
We show that binary mixtures of Bose-condensates of alkali atoms have a 
great variety of ground state and vortex structures which can be accessed
experimentally by varying the particle numbers of different alkalis. 
We have constructed a simple algorithm to determine the density profiles of
these states, and have worked out their phase diagrams within Thomas-Fermi
approximation. Many structures of the alkali binary contain a coexisting 
region, which is the analog of the long sought $^{3}$He-$^{4}$He 
interpenetrating superfluids in ultra-low temperature physics. 
\end{abstract}

The search of Bose condensate in alkali 
atoms\cite{Rb}\cite{Li}\cite{Na} has a deep root in ultra-low temperature 
physics. Since the discovery of superfluid $^{3}$He,  the searches of the next 
elemental superfluid have been focusing on
spin polarized hydrogen and $^{3}$He-$^{4}$He mixture.  The former promises 
another Bose superfluid besides the only known example of $^{4}$He, the latter,
the first example of interpenetrating superfluids.  
The recent discoveries of alkali Bose condensates\cite{Rb}\cite{Li}\cite{Na} 
have in essence achieved the goal of the superfluid hydrogen search. 
Since there are no intrinsic difficulties in loading more than one alkali
element and have them cooled in the same trap, it appears highly
promising that interpenetrating superfluids may be realized 
for the first time within the same experimental setting. 

In this paper, we shall discuss binary mixtures of alkali condensates. 
Such mixtures may consist of different alkalis such as 
$^{87}$Rb-$^{23}$Na, or different isotopes such as 
$^{87}$Rb-$^{85}$Rb, or different hyperfine states of the same alkali such
as the ($F=2, M_{F}=2$) and ($F=1, M_{F}=1$) states of $^{87}$Rb. 
We shall denote the two different alkalis as 1 and 2, 
and their particle numbers as $N_{1}$ and $N_{2}$. 
Unlike single component systems which are characterized by a single scattering
length, alkali binaries are  characterized by three scattering 
lengths $a_{1}, a_{2}$, and $a_{12}$; representing interactions between 
like and unlike alkalis. (While scattering lengths of like alkalis are
known at present, those of unlike alkalis are not). 
As we shall see, this moderate 
increase in energy scales leads to a proliferation of ground state and 
vortex structures, which 
we shall illustrate for the case $a_{1}, a_{2}>0$. This case is chosen 
because it has the greatest structural diversity, and can be analysed by 
simple analytic methods. The results for this
case will also be useful in understanding the qualitative features of other
(negative scattering length) cases. 

Our key results are : (a) a simple algorithm for determining the density 
profiles of the binary mixtures, (b) the phase diagrams of the vortex free 
ground states [denoted as (${\bf v0}$)], and the vortex states in either 1 or 2 
[denoted as (${\bf v1}$) and (${\bf v2}$) respectively].  We limit our
discussions to  these two types of vortices 
because they are the states the system will first 
fall into as the confining potential is rotated. Our algorithm, 
however, can be applied to arbitrary number of vortices in 1 and 2. 
As we shall see,  the structure of the mixture depends on two parameters
$\alpha$ and $\beta$, which are proportional to the strength of $a_{12}$ and 
the relative strength between $a_{1}$ and $a_{2}$ respectively. These
parameters determine whether alkali 2 when added to an existing cloud of 1 
will stay at its exterior or interior. 
Many structures of the mixture contain a coexisting region which is especially
large when $N_{1}\sim N_{2}$.  This is the analog of the 
long sought $^{3}$He-$^{4}$He superfluid mixture in ultra-low temperature 
physics. Even greater structural diversity is found when vortices are
inserted in one of the alkalis. When the alkali 1 and 2 are mutually repulsive, 
$a_{12}>0$, the vortex free alkali (say, 2) 
always enter the vortex core of 1, giving rise to a variety of 
``vortex donut" structures in 1. When $a_{12}<0$, one has an unusual 
``concentric donut" structure where a vortex free ``donut" of 2 embedded 
in a ``vortex donut" of 1. 

Our results are obtained by minimizing the Gross-Piteavskii energy 
$E(\Psi_{1}, \Psi_{2})=T+V$ subject to the constraint of constant particle 
numbers, i.e. by the condition $\delta K =0$, $K \equiv
E(\Psi_{1}, \Psi_{2})-\mu_{1}N_{1}-\mu_{2}N_{2}$, where 
($\Psi_{i}, \mu_{i}$) are the order parameter
and chemical potential of the $i$-th alkali, $i=1,2$. 
The potential energy $V$ and kinetic energy $T$ are 
\begin{equation}
V = \sum_{i=1,2}\int \left(
U_{i}({\bf x})\left|\Psi_{i}\right|^{2} +
\frac{2\pi\hbar^{2}a_{i}}{M_{i}}\left|\Psi_{i}\right|^{4}\right)
+ \int \frac{2\pi\hbar^{2}a_{12}}{\sqrt{M_{1}M_{2}}}\left|\Psi_{1}\right|^{2}
\left|\Psi_{2}\right|^{2} \label{V} \end{equation}
\begin{equation}
T= \int
\left(\sum_{i=1,2}\frac{\hbar^{2}\left|{\bf \nabla}\Psi_{i}\right|^{2}}{2M_{i}}
+ \zeta_{1} \Psi^{\ast}_{1}{\bf \nabla}\Psi_{1} \cdot
\Psi^{\ast}_{2}{\bf \nabla}\Psi_{2}
+\zeta_{2} \Psi^{\ast}_{1}{\bf \nabla}\Psi_{1} \cdot \Psi_{2}{\bf
\nabla}\Psi^{\ast}_{2}  + c.c. \right)  . \label{T} 
\end{equation}
$U_{i}$ is the potential of the magnetic trap for the $i$-th alkali, 
generally of the form\cite{HS}
$U_{i}({\bf x})$$=$
$\frac{g_{i}\mu_{B}B_{o}}{2L^{2}}$$(r^{2}+\lambda^{2}z^{2})$$\equiv$
$\frac{1}{2}M_{i}\omega_{i}^{2}$$(r^{2}+\lambda^{2}z^{2})$, 
where $g_{i}$ and $M_{i}$ are the g-factor and mass of the $i$-th alkali, 
$\mu_{B}$ is the Bohr magneton, $B_{o}$ is the magnetic field at the center
of the trap, $L$ is the length scale of the variations of the 
magnetic field, 
and $\lambda$ is the trap anisotropy.
$\zeta_{1}$ and $\zeta_{2}$ are complex coefficients which is caused by 
back flow effects between
different alkalis. They are expected to be small in the dilute limit.

Our calculations are performed within the Thomas-Fermi approximation (TFA), 
which is a good approximation in the limit of large number of particles, as 
pointed out by Baym and Pethick\cite{Baym}. In TFA, one ignores all 
${\bf \nabla}|\Psi_{i}|$ terms in $T$. For vortex free structures ${\bf (v0)}$,
this amounts to setting $T=0$. For ${\bf (v1)}$ vortices, 
($\Psi_{1}= |\Psi_{1}|e^{i\phi}$ and $\Psi_{2} = |\Psi_{2}|$),
TFA amounts to retaining only the 
centrifugal term $|\Psi_{1}|^{2}/r^{2}$ in $T$. Because of the absence of 
gradient terms ${\bf \nabla}|\Psi_{i}|$, the densities obtained from 
TFA usually consist of discontinuities in radial curvature. Such changes in 
curvature in fact take place smoothly over the distance of correlation length,
which decreases with increasing particle numbers\cite{Baym}. 
It is convenient to write 
$K$ and $N_{i}$ in dimensionless form by the following rescaling: 
(i) $z\rightarrow z/\lambda$, (ii) ${\bf x}\rightarrow
a_{T}{\bf x}$, where $\hbar^{2}/M_{1}a_{T}^{4}=g_{1}\mu_{B}B_{o}/L^{2}$,
(iii) $|\Psi_{i}|^{2}=A_{i}\rho_{i}$, where
$A_{1}=M_{1}\omega_{1}/(8\pi\hbar a_{1})$,
$A_{2}/A_{1} = (a_{1}M_{2}/a_{2}M_{1})^{1/2}$,
(iv) $\mu_{1}\equiv \frac{1}{2}\hbar\omega_{1}\nu_{1}$
and $\mu_{2}\equiv \frac{1}{2}\hbar\omega_{1}(A_{2}/A_{1})\nu_{2}$,
(v) $K\equiv(\hbar\omega_{1}/4\lambda)(a_{T}/(4\pi a_{1}){\cal K}$.
(vi) $N_{1}=(8\pi\lambda)^{-1}(a_{T}/a_{1})n_{1}$,
$N_{2}=(8\pi\lambda)^{-1}(a_{T}/a_{1})\sqrt{a_{1}M_{2}/a_{2}M_{1}}n_{2}$.
We then have 
\begin{equation}
{\cal K} =  \int {\rm d}^{3}x  \left[ -c_{1}\rho_{1}-c_{2}\rho_{2} +
\frac{1}{2}\left(\rho_{1}^{2} + \rho_{2}^{2} +
2\alpha\rho_{1}\rho_{2}\right)\right] ,
\,\,\,\,\,\,\,\,\, n_{i} = \int {\rm d}^{3}x \rho_{i}
\label{kappa}
\end{equation}
\begin{equation}
c_{1}({\bf x}) = \nu_{1}-\left( r^{2} + \frac{p}{r^{2}} +z^{2}\right) 
\,\,\,\,\,\,
c_{2}({\bf x}) = \nu_{2} - \beta (r^{2}+z^{2}) , 
\label{c1} \end{equation}
where ${\bf x}=({\bf r},z)=(x,y,z)$ in eq.(\ref{c1}), and 
$p=0$ and 1 for ${\bf (v0)}$ and ${\bf (v1)}$ state respectively.  
$\alpha$ and $\beta$ are defined as
\begin{equation}
\alpha = \frac{a_{12}}{2\sqrt{a_{1}a_{2}}}, \,\,\,\,\,\,\
\beta = \frac{g_{2}}{g_{1}}\sqrt{\frac{a_{1}M_{2}}{a_{2}M_{1}}} , 
\label{alphabeta} \end{equation}
reflecting the interaction between unlike alkalis, and 
the relative strength of the self interaction between like alkalis.
For ${\bf (v2)}$ vortices, the energy can be rescaled to the same form as 
eq.(\ref{kappa}) with the same $\alpha$ and $\beta$ in eq.(\ref{alphabeta}), 
and with ($c_{1}$$=$$r^{2}+z^{2}-$$\nu_{1}$,
$c_{2}$$=$$\beta$$(r^{-2}+r^{2}+z^{2})-$$\nu_{2}$) \cite{v2}. 
We shall not consider the case $(\alpha<-1)$, which does not have bulk
stability.  In our subsequent illustrations, we use 
the example of $^{87}$Rb-$^{23}$Na mixture, (our alkali 1 and
2), which has identical $g$ factors. We thus have 
$a_{1}\approx 100\AA$\cite{Rb}, $a_{2}\approx 49\AA$\cite{Na}, and $\beta=0.73$. 
Since $a_{12}$ is not known for 
this mixture, we consider the entire range of $\alpha >-1$. 
Taking $\lambda=1$ and
$a_{T}=4\times 10^{-4}$cm, we have $N_{1}=16.7 n_{1}, N_{2}=12.3 n_{2}$. 

Our goal is to find the condensate structure (i.e. $\rho_{1}, \rho_{2}$) 
as a function of particle number $n_{1}, n_{2}$. This is done by : {\bf (I)}
Minimizing ${\cal K}$ to find the equilibrium densities $\rho_{i}$ for given 
chemical potential $\nu_{i}$, 
{\bf (II)} substituting these densities into eq.(\ref{kappa}) to obtain the 
relation $n_{i}=n_{i}(\nu_{1}, \nu_{2})$, 
{\bf (III)} inverting this relation to obtain $\nu_{i}=\nu_{i}(n_{1}, n_{2})$ 
and hence the evolution of $\rho_{i}$ (through their dependence
on $\nu_{i}$) as a function of $n_{i}$. Although much of our labor went into
${\bf (II)}$ and ${\bf (III)}$, 
they are straightforward (though lengthy) calculations once the densities 
profiles are determined by the simpler but subtler step ${\bf (I)}$, 
which we now explain. 

Let ${\bf [0]}, {\bf [1]}, {\bf [2]}, {\bf [12]}$ denote the vacuum, the 
single phase of 1, 2, and the coexisting phase of 1 and 2 respectively. 
Their densities are given by the stationary 
conditions of ${\cal K}$, 
\begin{equation}
{\bf [12] :} \,\,\,\,\, 
\rho_{1}=\frac{c_{1} -\alpha c_{2}}{1-\alpha^{2}} \geq 0 , \,\,\,\,\,\,
\rho_{2}=\frac{c_{2} -\alpha c_{1}}{1-\alpha^{2}} \geq 0 ; \hspace{0.5in}
{\bf [0] :} \,\,\,\,\, \rho_{1}=\rho_{2}=0 ; 
\label{onetwo} \end{equation}
\begin{equation}
{\bf [1] :} \,\,\,\,\, \rho_{1}=c_{1} \geq 0,  \,\,\,  \rho_{2}=0, 
\,\,\,\,\,\,\,\,\,\,\,\,\,\,\,
{\bf [2] :} \,\,\,\,\, \rho_{2}=c_{2} \geq 0,  \,\,\,  \rho_{1}=0. 
\label{one} \end{equation}
The distribution of these ``phases" in $c_{1}$-$c_{2}$ plane will be 
referred to the {\em distribution plot}, which depends only on $\alpha$.
The distribution plots for ($0<\alpha<1$) and ($-1<\alpha<0$)
are shown in fig.(1.1) and fig.(1.3). [The distribution plot for 
($\alpha >1$) will not be shown as it is given by fig.(1.1) with the coexisting
region ${\bf [12]}$ collapsed into the line $c_{1}=c_{2}$.]
The boundaries separating ${\bf [12]}$ from ${\bf [2]}$ and ${\bf [1]}$
are denoted as ${\bf 1_{o}}$ and ${\bf 2_{o}}$.
They are the surfaces of vanishing $\rho_{1}$ and $\rho_{2}$,  described by
equations (${\bf 1_{o} :}$ $c_{1}=\alpha c_{2}$) and  (${\bf 2_{o} :}$
$c_{2}=\alpha c_{1}$) respectively. 
The boundaries separating ${\bf [0]}$ from ${\bf [1]}$ and ${\bf [2]}$ 
are denoted as ${\bf 1^{o}}$ and ${\bf 2^{o}}$. They are described by equations 
(${\bf 1^{o} :}$ $c_{1}=0$) and  (${\bf 2^{o} :}$ $c_{2}=0$). 
>From eq.(\ref{c1}), one can see that a path in real space will have an 
``image" path in $c_{1}$-$c_{2}$ space. For example, the image of a radial path 
on the horizontal plane with height $z$ is given by 
$\Gamma(z) : c_{1}-\nu_{1}(z)$$=$
$\beta^{-1}[c_{2}-\nu_{2}(z)]+p\beta[c_{2}-\nu_{2}(z)]^{-1}$, where
$\nu_{1}(z)=\nu_{1}-z^{2}$, $\nu_{2}(z)=\nu_{2}-\beta z^{2}$. 
For vortex free state ($p=0$), $\Gamma(z)$ is a straight line with
slope $\beta$ emerging from the point  ($\nu_{1}(z), \nu_{2}(z)$), which is
shown as a dashed line emerging from a circle in fig.(1.1). 
For the ${\bf (v1)}$ vortices, $\{ \Gamma(z)\}$ is a family of curves shown in
fig.(1.2), where $z_{2}>z_{1}>0$ in this figure. 
The arrows on these paths indicate the direction of increasing
$r$. As $r$ varies from 0 to $\infty$, $\Gamma(z)$ 
intersects the phase boundaries in a specific order. 
>From the definition of $\nu_{i}(z)$, it is easy to see that as 
$|z|$ increases, $\Gamma(z)$ slides down rigidly along
the straight line with slope $\beta$. 

The above considerations suggest a simple algorithm for determining the
structure of the alkali binary : (i) For given $\nu_{1}, \nu_{2}$, draw the
image paths $\{ \Gamma(z) \}$ on the distribution plot. $\Gamma(z)$ will
intersect a set of phase boundaries in a specific order. (ii) Construct the set
of boundary surfaces intersected by $\{ \Gamma(z)\}$.
(iii) Eliminate all portions of the boundary surfaces that are 
inconsistent with the order of intersections generated in (i).  The 
remaining surfaces are the physical boundaries in the mixture.
The densities of various phases bounded by these surfaces are given by
eq.(\ref{onetwo}) and (\ref{one}).

To illustrate this algorithm, consider the vortex free mixture in fig.(1.1). 
The image paths of this mixture intersect phase boundaries 
${\bf 2_{o}, 1_{o}}$ and ${\bf 2^{o}}$, corresponding
to a set of concentric spherical surfaces in real space separating 
the single component and  coexisting regions as shown in fig.(1.4). Fig.(1.4)
is the structure of the mixture. 
For the ${\bf (v1)}$ vortex in fig.(1.2), the family $\{ \Gamma (z)\}$ 
intersects all four phase boundaries 
(${\bf 1_{o}}$, ${\bf 2_{o}}$, ${\bf 1^{o}}$, ${\bf 2^{o}}$) 
in the order (${\bf 1_{o}2_{o}1^{o}}$), or ${\bf 2^{o}}$ alone. 
Displaying all four boundary surfaces as in fig.(1.5) and eliminating all 
sections of these surfaces inconsistent with 
the order of intersection, we obtain the 
physical boundaries shown in fig.(1.6). 
The densities of the phases ${\bf [1]}$, ${\bf [12]}$, and ${\bf [2]}$ 
enclosed by these surfaces are given in eq.(\ref{onetwo}) and (\ref{one}). 
It is clear that this algorithm applies to mixtures with arbitrary 
number of vortices in 1 and 2, which have different image trajectories. 

Having determined the density profiles for given chemical potential $(\nu_{1},
\nu_{2})$, we have followed the aforementioned 
Steps $({\bf II})$ and $({\bf III})$ to construct the 
phase diagrams for both ground states [${\bf (v0)}$] and vortex states 
[${\bf (v1)}$ and ${\bf (v2)}$] over the entire range of ($\alpha,\beta$).
For reasons to be explained shortly, we shall focus on the parameter range 
[{\bf A}: $0<\alpha<1$, $\beta<\alpha$],
[{\bf B}: $0<\alpha<1$, $\alpha<\beta<1$],
and [{\bf C}: $-1<\alpha<0, \beta<1$].
For brevity, we shall denote the ${\bf (v0)}$ state of ${\bf A}$ as 
${\bf (v0)}$-${\bf A}$, the ${\bf (v2)}$ state of ${\bf B}$ as
${\bf (v2)}$-${\bf B}$, etc. As it turns out, for ground state phase diagrams,
it is sufficient to discuss those of ${\bf (v0)}$-${\bf A}$ and 
${\bf (v0)}$-${\bf B}$. For vortex phase diagrams, 
it is sufficient to discuss those of ${\bf (v1)}$-${\bf A}$,
${\bf (v2)}$-${\bf A}$, ${\bf (v1)}$-${\bf B}$, and ${\bf (v1)}$-${\bf C}$. 
The phase diagrams of all other cases can be obtained from these either 
by interchanging 1 and 2 or by collapsing the coexistence region down 
to a line\cite{comment1}. 
In all cases we have studied, the condensate structure evolves $continuously$
 over the entire $n_{1}$-$n_{2}$ space, even though the phase diagrams may
contain different structural regimes. 

The typical phase diagrams of ${\bf (v0)}$-${\bf A}$ and  
${\bf (v0)}$-${\bf B}$  are shown in fig.(2.1) and (2.2), with 
($\alpha=0.9, \beta=0.73$) and  ($\alpha=0.6, \beta=0.73$) respectively.
The density profiles for mixtures marked by letters
``$a$" to ``$g$" in fig.(2.1) and (2.2) are shown in fig.(2.3) to (2.9). 
The line between $b$ and $c$ in fig.(2.1) divides the 
configurations where 2 is absent or present at the origin. 
The main difference between {\bf A} and {\bf B} 
is that the former ($\beta<\alpha$) represents a regime where repulsion 
between unlike alkalis dominates. As a result,
when a small amount of 2 is added to an existing condensate of 1,  it stays
at its surface in the case of {\bf A} (see fig.(2.3) and (2.4)), 
whereas it enters directly into the center of 1 in the case of {\bf B} (see 
fig.(2.7)).  One can see from fig.(2.5) and (2.8) that when 1 and 2 have
similar number of particles, the coexisting region of 1 and 2 occupies 
a substantial portion of the mixture. This large coexisting region is 
the analog of the long sought interpenetrating superfluid phase 
in $^{3}$He-$^{4}$He mixture.

Next, we turn to the vortex states. The presence of a vortex (say, in alkali 1) 
will turn the spherical ground state condensate into a vortex ``donut". The 
donut hole is the vortex core, around which supercurrent circulates. 
For repulsive interactions, ($\alpha >0$), 
the general behavior of the (vortex free) alkali 2 is to fill up the vortex
core and to penetrate into the interior of vortex 1. 
The typical phase diagram of ${\bf (v1)}$-${\bf A}$, ${\bf (v2)}$-${\bf A}$, 
and ${\bf (v1)}$-${\bf B}$
are shown in fig.(3.1), (3.5), and (3.7) respectively, with values of
$(\alpha=0.9, \beta=0.73)$ for ${\bf A}$ and 
$(\alpha=0.6, \beta=0.73)$ for ${\bf B}$. The scales of
$n_{1}$ and $n_{2}$ are chosen so that a condensate with 
$N_{1}=N_{2}=10^{4}$ will
appear in the middle of the diagram. There are in fact many more 
structural regimes near the $n_{1}$ and $n_{2}$ axes which correspond to
various stages of filling up the vortex core. However, they only exist close
to the $n_{1}$ and $n_{2}$ axes (i.e. small number of particles in either 1 or
2) and are not visible on this scale\cite{HS2}. 

The boundary surfaces for the mixtures 
marked as ``$a$" to ``$f$" in fig.(3.1), (3.5) and (3.7) are shown in the 
rest of figure 3 with self explanatory labelling. These surfaces are 
presented only in a quadrant of $r$-$z$ plane, as they are cylindrical 
symmetric about $z$ and have mirror symmetry about the $xy$ plane. 
The boundaries ${\bf 1_{o}}$ and 
 ${\bf 2_{o}}$ are represented by solid lines, while 
(${\bf 1^{o}}$ and ${\bf 2^{o}}$)  are represented by dashed lines.
1 and 2 coexist in the region between the dashed lines. 
The curve in fig.(3.1) between $a$ and $b$ divides regions where a ring of
single phase ${\bf [1]}$ is present or absent in the interior of 
the ``vortex donut".
The curve between $b$ and $c$ divides the regions where the
``vortex donut"  resides entirely inside or extends beyond the 
spherical cloud of 2. The phase diagram of ${\bf (v2)}$-${\bf A}$ is very 
simple as all mixture states have the same qualitative structure 
shown in fig.(3.6).
The phase diagram of ${\bf (v1)}$-${\bf B}$ (fig.(3.7)) is essentially that of 
${\bf (v1)}$-${\bf A}$ 
(fig.(3.1)) with the region occupied by ``$b$" is collapsed
into a line. The absence of this region is due to the weakness of mutual
repulsion between different alkalis (i.e. small $\alpha$), so that alkali 2 
always penetrate completely the vortex donut of 1. 

Finally, we consider the ${\bf (v1)}$-${\bf C}$ vortices, i.e. the vortex state
of a mixture with $a_{12}<0$, (hence $\alpha<0$). 
The typical phase diagram is shown in  fig.(4.1), (where we 
have chosen $\alpha=-0.9, \beta=0.73$). It is qualitatively the same for all
values of $\beta$. The boundary surfaces of the mixtures marked as 
$a$, $b$, and $c$ are shown in fig.(4.2) to (4.4).  The region in fig.(4.1) 
containing ``$a$"  describes the unusual structure of a vortex free 
donut of 2 is embedded in the vortex donut of 1. When
crossing the boundary separating $a$ and $b$ in fig.(4.1), the hole of 
``donut" 2 shrinks to zero, filling up the vortex core of 1. 
As one crosses the boundary separating 
$b$ and $c$ in fig.(4.1), the ``vortex donut" 1 is completely devoured by 2. 

>From the above discussions, we see that binary mixtures of alkali Bose
condensates possess a great variety of ground state and vortex
structures, which can be accessed by varying the number of particles of each 
alkali. 
This array of structures allows one the go continuously from a regime of 
interpenetrating superfluids to one with separated phases. The possibility of
scanning through this continuum offers great opportunities to study
coupled macroscopic quantum phenomena and interactions of 
elementary excitations in distinct Bose fluids. 
Realizations of alkali condensate mixtures will certainly deepen our
understanding of interacting Bose fluids, and widen our horizon on 
superfluid phenomenon.

TLH would like to thank Greg Lafyatis for discussions. This work is supported
in part by NSF Grant No. DMR-9406936.

\newpage

\noindent {\bf Caption}

\vspace{0.2in}

\noindent {\bf Figure 1:} Fig.(1.1) and (1.2) show the typical 
distribution plots for $0<\alpha<1$. The 
``image trajectories" $\Gamma(z)$ (dashed lines) of the ground state 
${\bf (v0)}$ and the ${\bf (v1)}$ vortex of a mixture with chemical potential 
$(\nu_{1}, \nu_{2})$ are shown as dashed lines in fig.(1.1) and (1.2).
In fig.(1.2), $z_{2}>z_{1}>0$.   Fig.(1.3) shows the distribution plot for 
$-1<\alpha<0$. 
Fig.(1.4) and fig.(1.5) to (1.6) show the real space boundary surfaces for
${\bf (v0)}$ and ${\bf (v1)}$ cases respectively.
The boundary surfaces in fig.(1.4) are given by 
$[{\bf 2^{o}:\,\,} r^{2}+z^{2}=\nu_{2}/\beta]$, 
$[{\bf 1_{o}:\,\,} r^{2}+z^{2}=(\nu_{1}-\alpha\nu_{2})/(1-\alpha\beta)]$, 
$[{\bf 2_{o}:\,\,} r^{2}+z^{2}=(\alpha\nu_{1}-\nu_{2})/(\alpha-\beta)]$. 
The boundary surfaces in fig.(1.5) are given by
$[{\bf 1^{o}:\,\,} r^{2}+r^{-2} +z^{2} = \nu_{1}]$, 
$[{\bf 2^{o}:\,\,} r^{2}+z^{2} = \nu_{2}/\beta]$,
$[{\bf 1_{o}:\,\,} (1-\alpha\beta)(r^{2}+z^{2})+r^{-2}=\nu_{1}-\alpha\nu_{2}]$,
$[{\bf 2_{o}:\,\,} (\alpha-\beta)(r^{2}+z^{2})+\alpha/r^{2}=
\alpha\nu_{1}-\nu_{2}]$. Fig.(1.6) gives the structure of the mixture in the 
$r$-$z$ plane. 

\noindent {\bf Figure 2:} Typical ground state phase diagrams for 
${\bf (v0)}$-${\bf A}$ and ${\bf (v0)}$-${\bf B}$. We
have chosen $(\alpha=0.9, \beta=0.73)$ for ${\bf A}$, 
$(\alpha=0.6, \beta=0.73)$ for ${\bf B}$. The dotted line in fig.(2.1)
represents the case of equal particle numbers in both condensates,
($N_{1}=N_{2}$). As discussed in the text, for parameters appropriate for 
$^{87}$Rb and $^{23}$Na, and for current magnetic trap, we have 
$N_{1}=16.7 n_{1}, N_{2}=12.3 n_{2}$. The circle in fig.(2.1) indicate 
a mixture with $N_{1}=N_{2}=10^{4}$. 
For the densities shown in fig.(2.3) to (2.9), the chemical potential $\nu$ for
the outer cloud is of the order of 10 to 20, giving rise to a mixture size of
about 3 to 4 in dimensionless units, which is about $2\times 10^{-3}$cm for 
a cloud of $N_{1}\approx 10^{4}$, $N_{2}\approx 10^{4}$.

\noindent {\bf Figure 3:} Typical phase diagrams for 
${\bf (v1)}$ and ${\bf (v2)}$
vortices for parameter range [${\bf A} : 0<\alpha<1, \beta<\alpha$], 
and [${\bf B} : 0<\alpha<1, \alpha<\beta$]. The values of $(\alpha, \beta)$
used here are identical to those used in figure 2. 
Figures with letter labels ``$a$", 
``$b$", etc. at the upper right corners show the boundary surfaces of the
mixtures indicated by the same letter in the $n_{1}$-$n_{2}$ phase
diagrams. 

\noindent {\bf Figure 4:} Fig.(4.1) shows the typical phase diagram 
for the ${\bf (v1)}$ vortices for case [${\bf C}$: $\alpha <0$], with 
$(\alpha=-0.9, \beta=0.73)$. The notations here are identical
to those in figure 3. 


\begin{references}
\bibitem{comment1}   All cases with $\beta>1$ can be obtained from those of
$\beta<1$ by interchanging 1 and 2. All cases with $\alpha>1$ can be obtained
from those with  $\alpha<1$ by shrinking the coexistence phase down to a line. 
The phase diagram of ${\bf (v0)}$-${\bf C}$ is qualitatively the same 
as ${\bf (v0)}$-${\bf B}$, (i.e. fig.(2.2)) and need not be discussed
separately. 
Similarly, ${\bf (v2)}$-${\bf B}$ and  ${\bf (v2)}$-${\bf C}$ are qualitatively
similar to ${\bf (v1)}$-${\bf B}$ and  ${\bf (v1)}$-${\bf C}$, with $n_{1}$ and
$n_{2}$ axes interchanged.
\end{references}
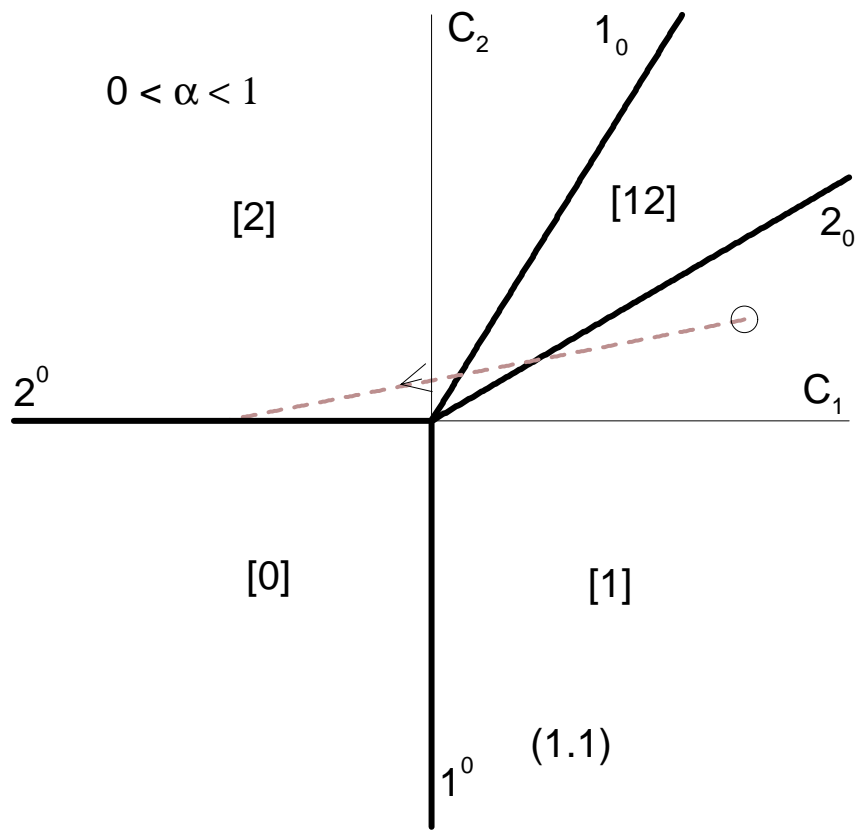
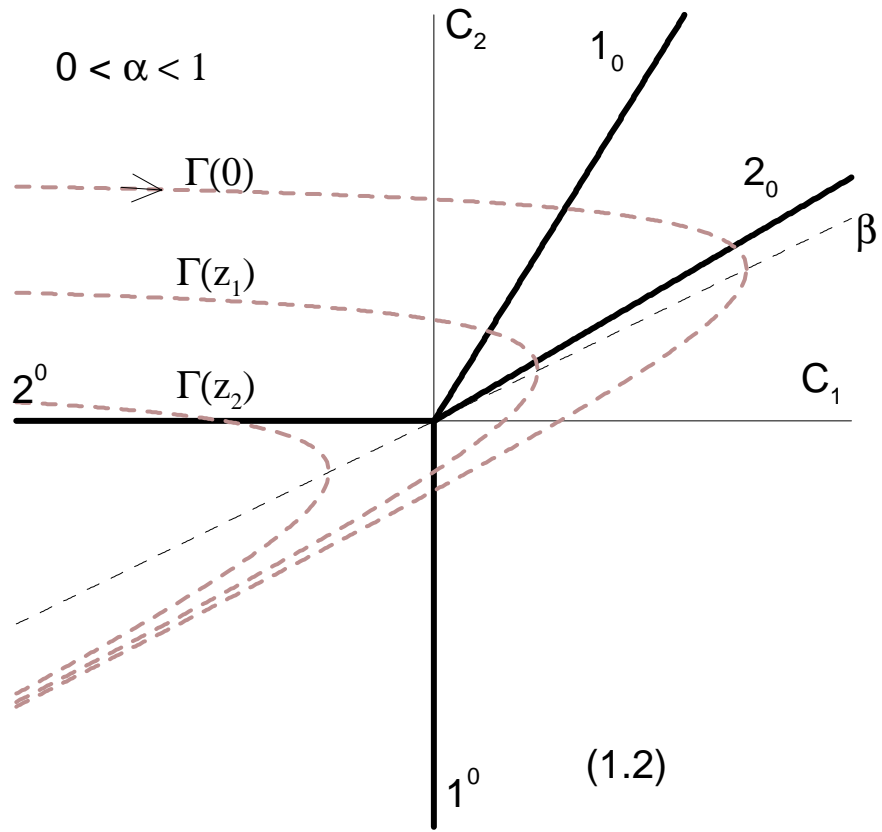
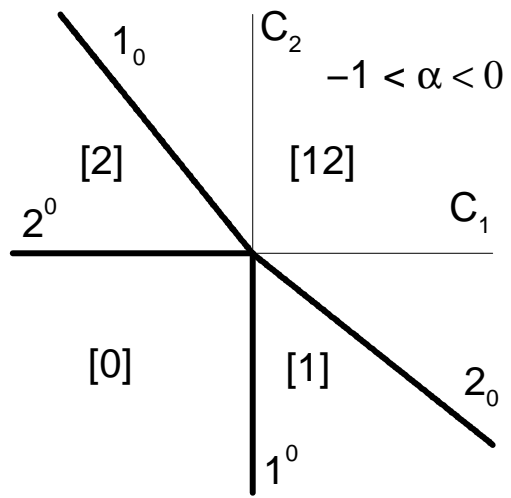
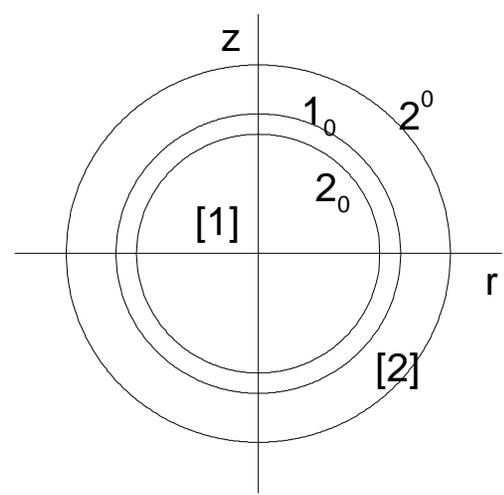
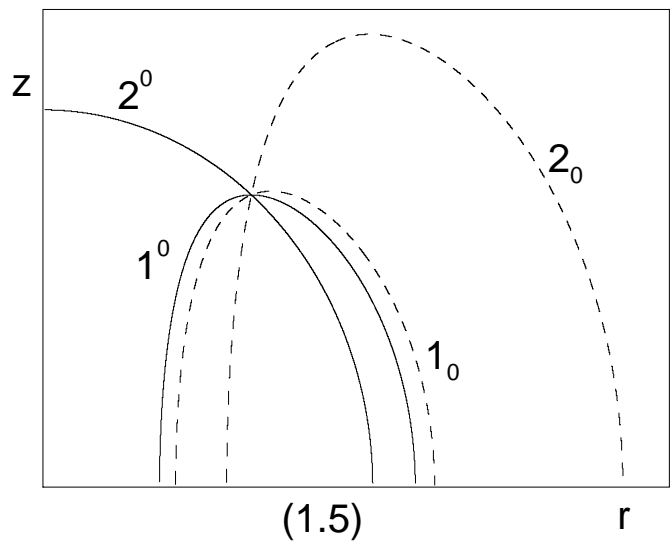
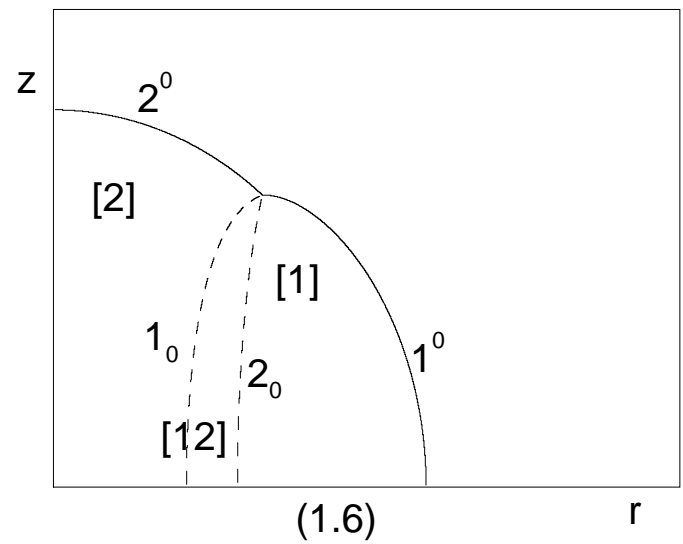
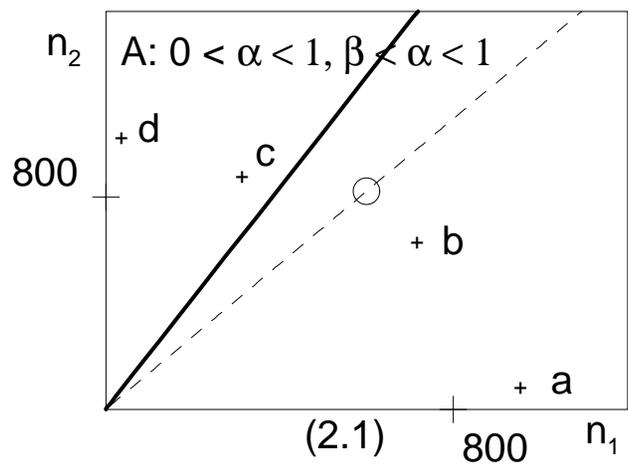
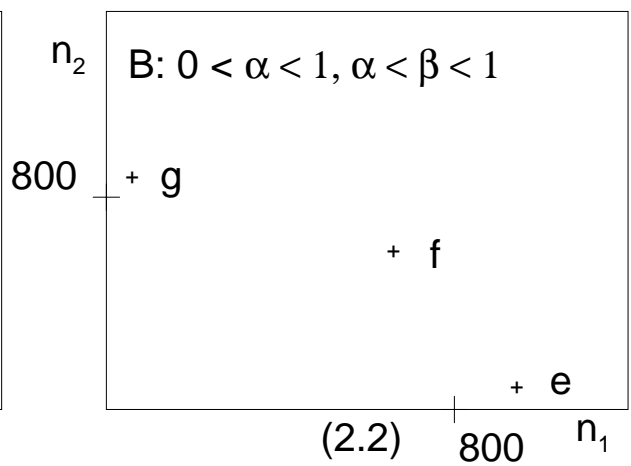
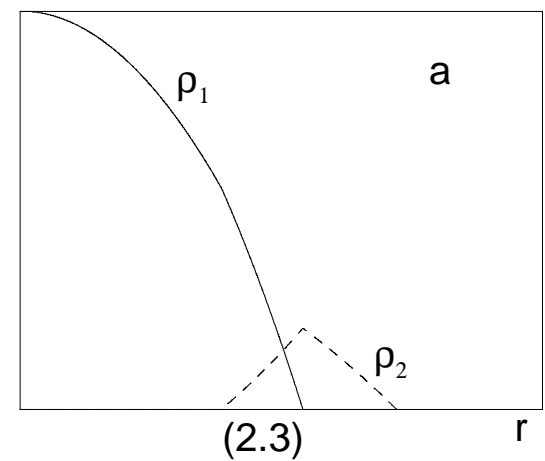
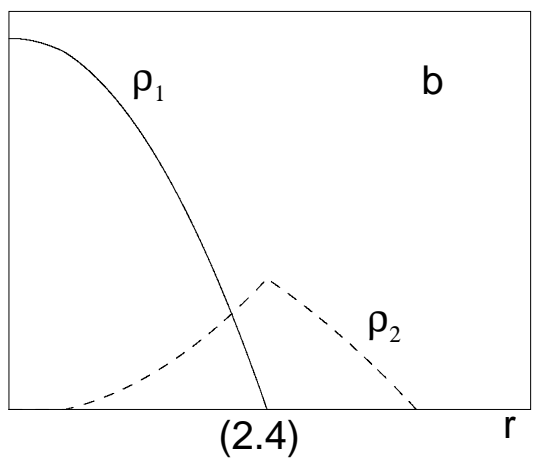
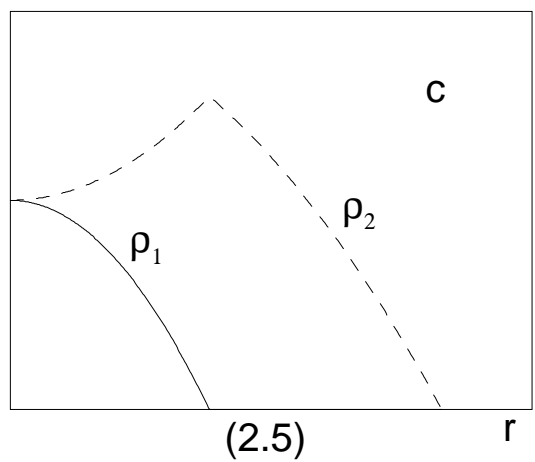
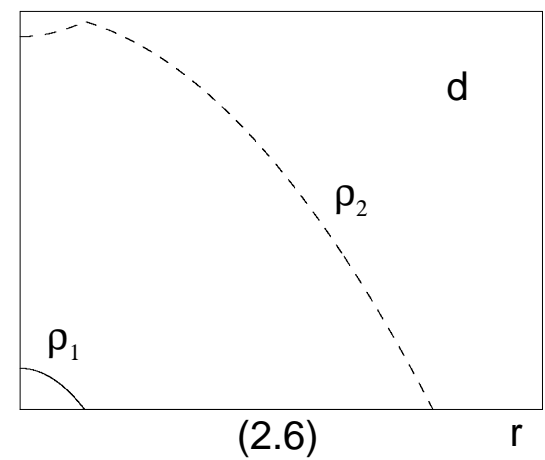
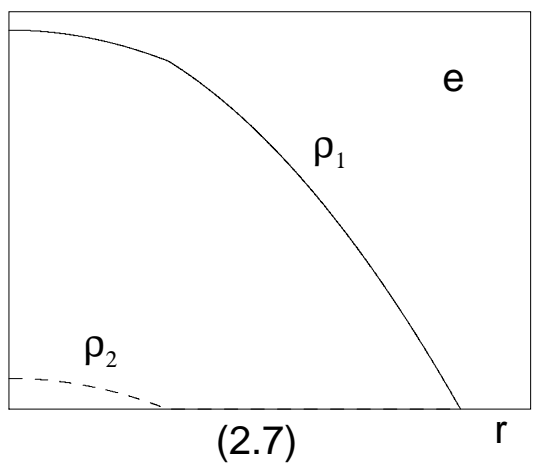
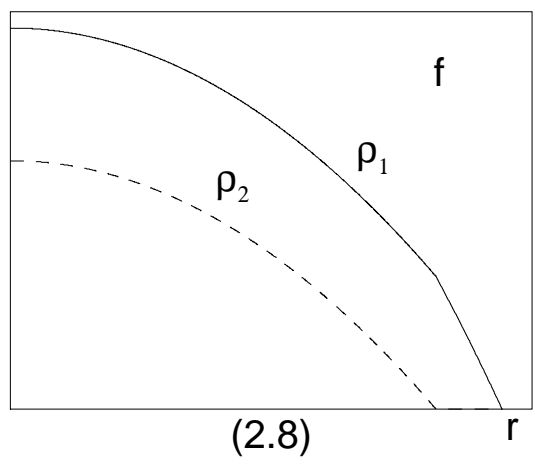
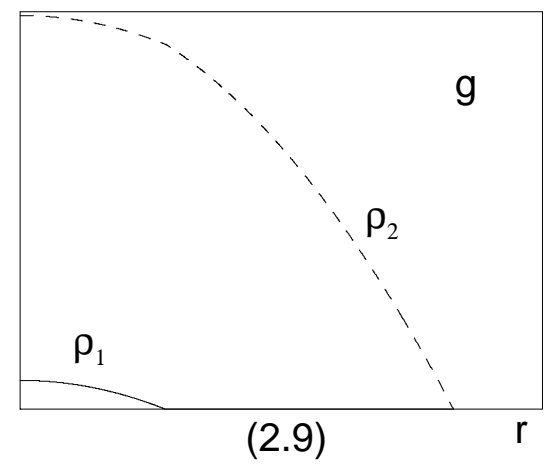
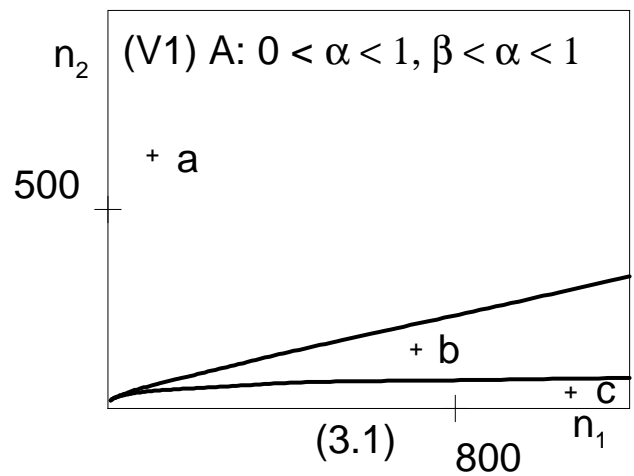
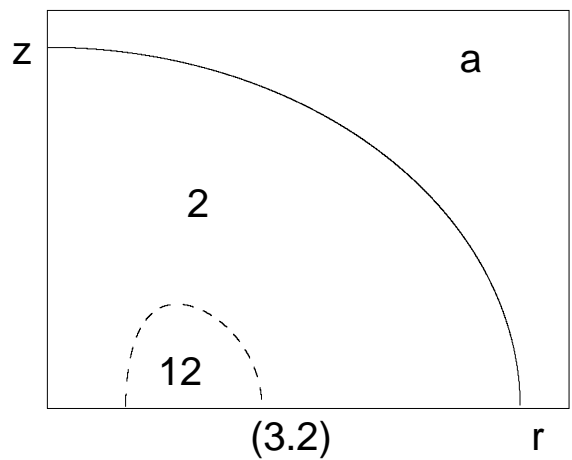
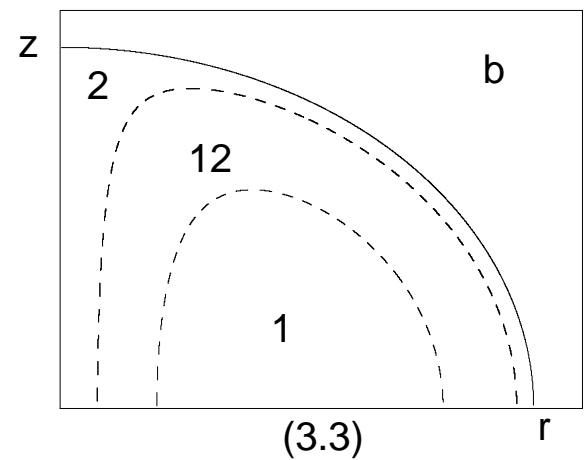
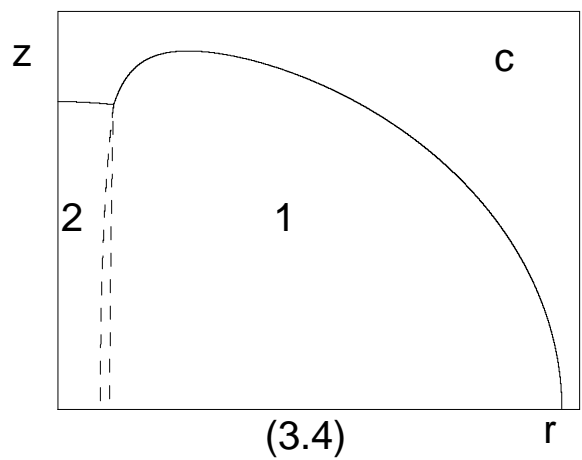
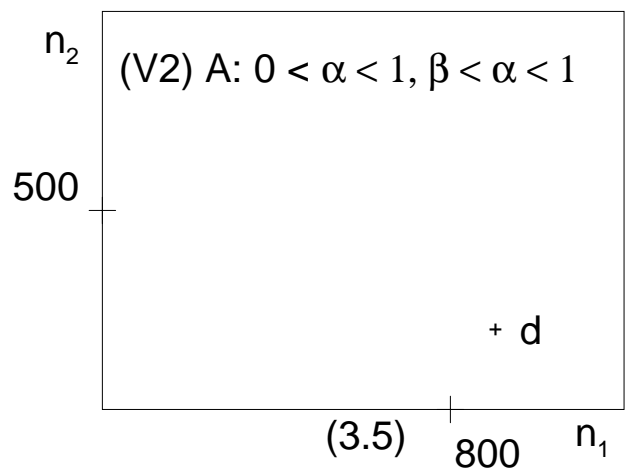
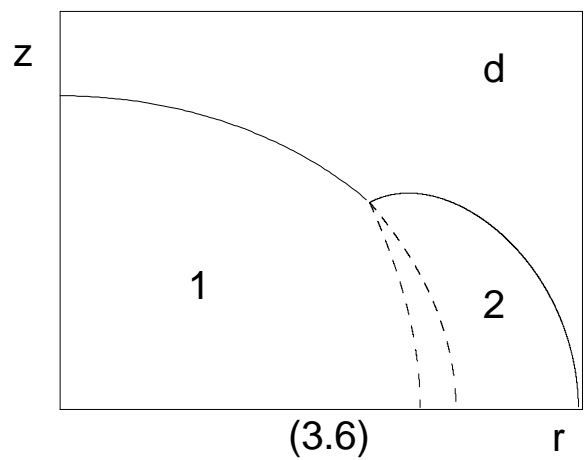
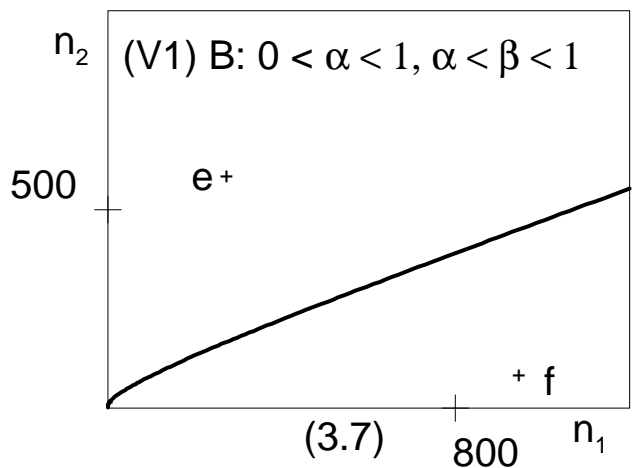
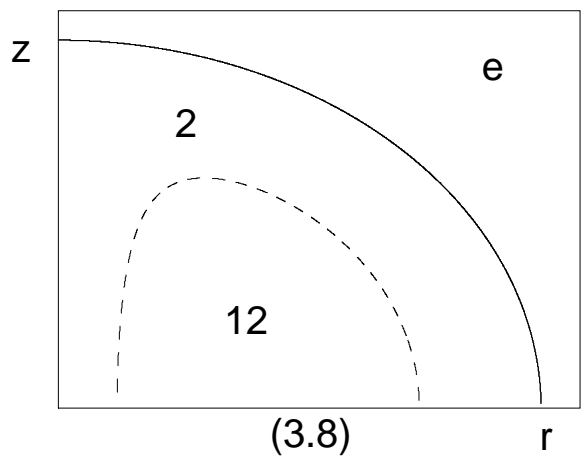
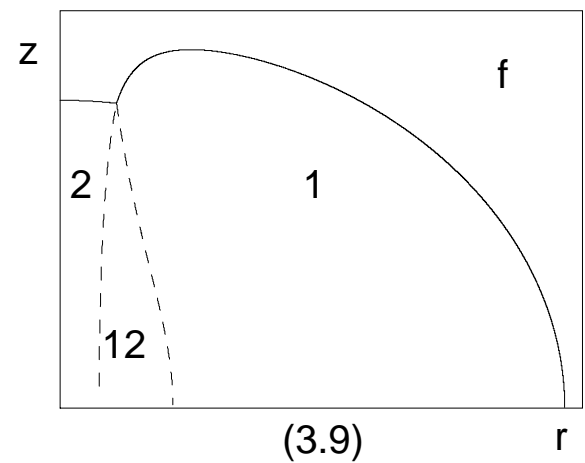
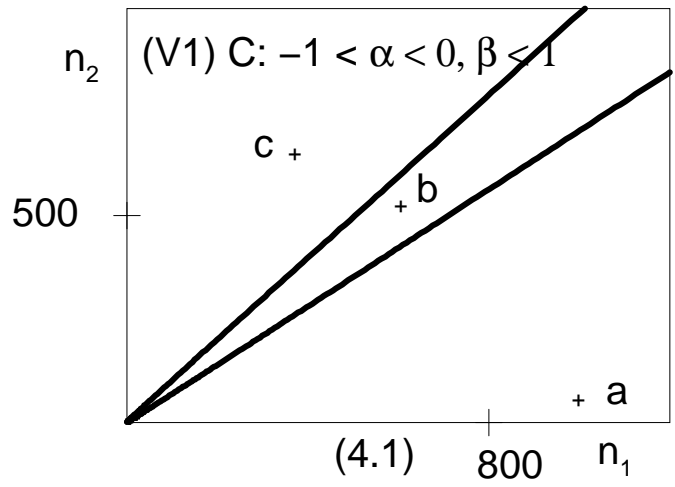
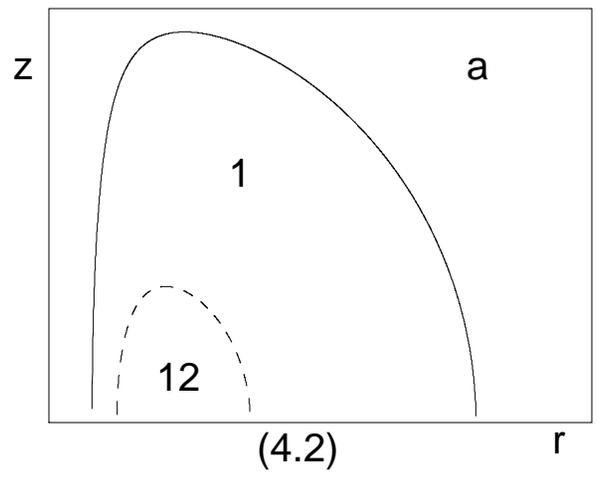
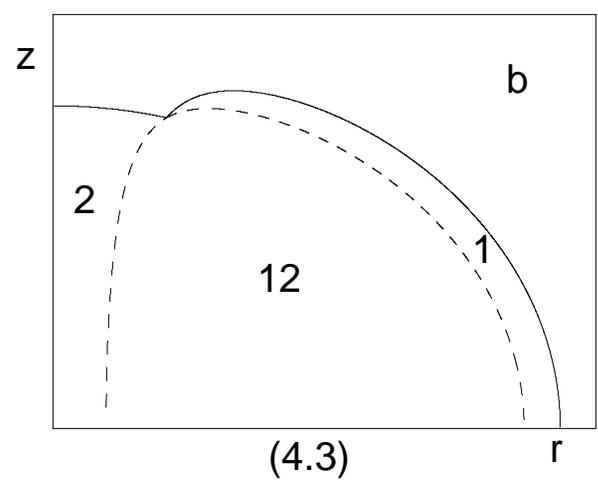
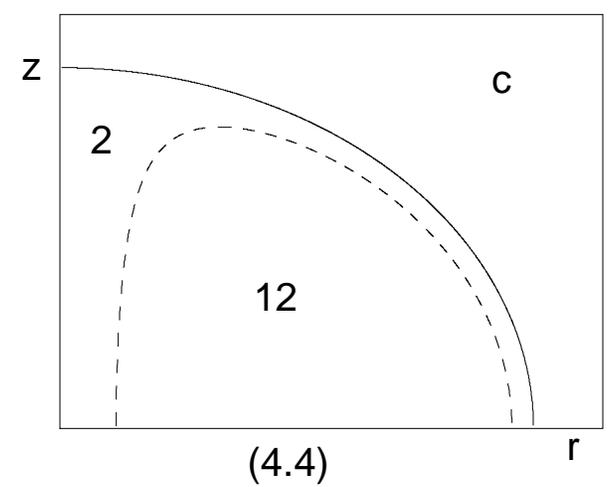
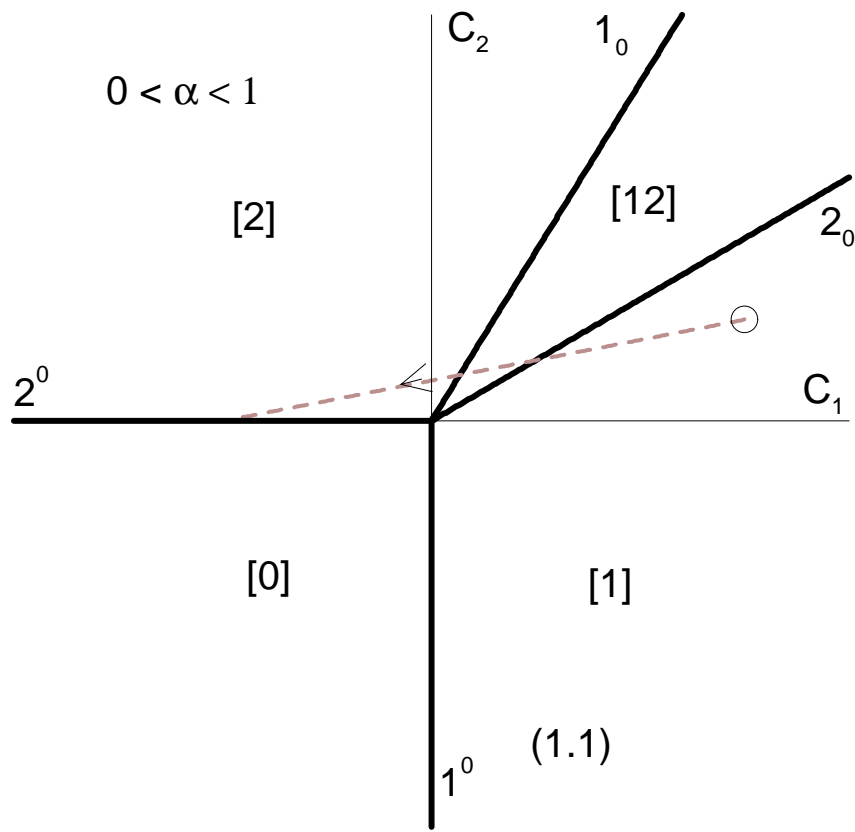
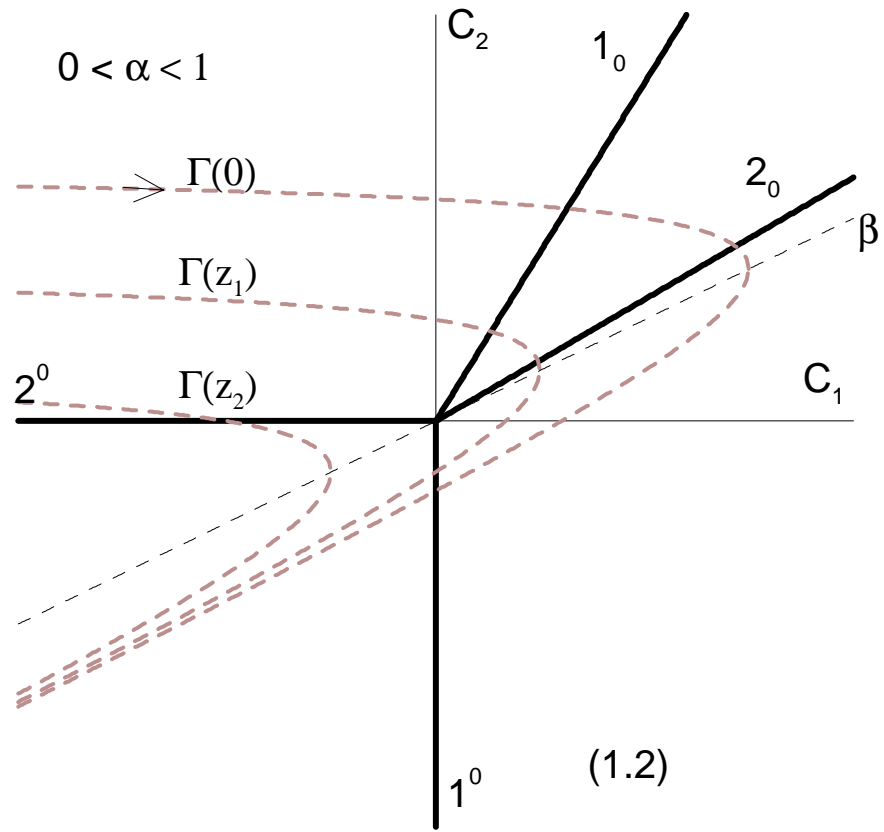
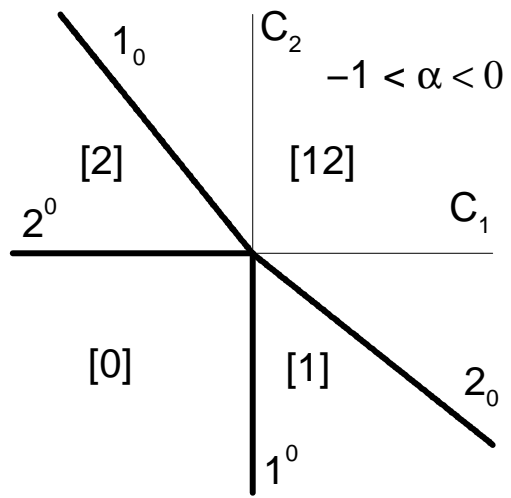
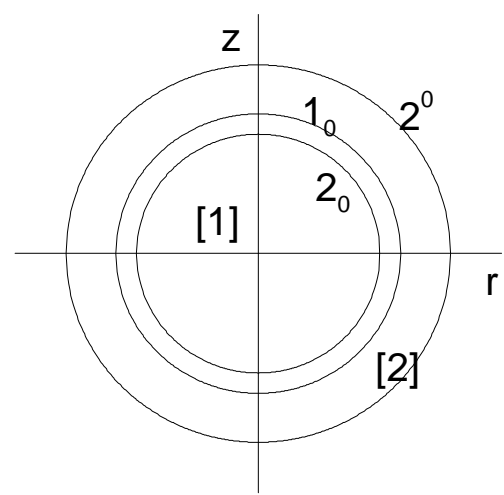
\end{document}